\documentclass[prb,twocolumn,floatfix,showpacs]{revtex4-1}
\usepackage{graphicx,color}
\usepackage{amsmath,amssymb,bm}

\newcommand{\tn}{\textnormal}
\newcommand{\cprb}[3]{Phys.~Rev.~B {\bf #1}, #2 (#3)}
\newcommand{\cprl}[3]{Phys.~Rev.~Lett.~{\bf #1}, #2 (#3)}

\definecolor{darkred}{rgb}{0.90,0,0}
\definecolor{darkgreen}{rgb}{0,0.60,.2}
\definecolor{darkblue}{rgb}{0,0,1}
\definecolor{grey}{cmyk}{0,0,0,0.25}
\definecolor{orange}{cmyk}{0,0.6,0.8,0}

\begin{document}
\title{\boldmath Luttinger liquid physics from infinite-system DMRG}
\author{C.\ Karrasch$^{1}$}
\author{J.\ E.\ Moore$^{1,2}$}

\affiliation{$^1$Department of Physics, University of California, Berkeley, California 95720, USA}

\affiliation{$^2$Materials Sciences Division, Lawrence Berkeley National Laboratory, Berkeley, CA 94720, USA}

\begin{abstract}

We study one-dimensional spinless fermions at zero and finite temperature $T$ using the density matrix renormalization group. We consider nearest as well as next-nearest neighbor interactions; the latter render the system inaccessible by a Bethe ansatz treatment. Using an infinite-system alogrithm we demonstrate the emergence of Luttinger liquid physics at low energies for a variety of static correlation functions as well as for thermodynamic properties. The characteristic power law suppression of the momentum distribution $n(k)$ function at $T=0$ can be directly observed over several orders of magnitude. At finite temperature, we show that $n(k)$ obeys a scaling relation. The Luttinger liquid parameter and the renormalized Fermi velocity can be extracted from the density response function, the specific heat, and/or the susceptibility without the need to carry out any finite-size analysis. We illustrate that the energy scale below which Luttinger liquid power laws manifest vanishes as the half-filled system is driven into a gapped phase by large interactions.

\end{abstract}

\maketitle

% 71.10.Pm  Fermions in reduced dimensions
% 71.10.Fd  Lattice fermion models
% 71.27.+a  Strongly correlated electron systems

%%%%%%%%%%%%%%%%%%%%%%%%%%%%%%%%%%%%%%%%   Introduction

\section{Introduction}

The concept of a Luttinger liquid (LL) is one of the most striking manifestations of strong electronic interactions in condensed matter theory.\cite{voit,giamarchi,kurtrev1,kurtrev2} It asserts that the low-energy excitations of a large class of gapless one-dimensional systems are not fermionic quasiparticles but collective (bosonic) modes; as a consequence, correlation functions exhibit anomalous power laws in space and time with interaction-dependent exponents. It is one of the hallmarks of a Luttinger liquid that the fermionic momentum distribution function $n(k)$ exhibits a zero-temperature power law singularity near the Fermi momentum instead of a Fermi liquid discontinuity.

The Luttinger liquid phenomenology is usually established by a two-step protocol. One first shows that under some rather \textit{restrictive assumptions} -- most importantly of a linear dispersion $\pm v_F^0|k-k_F|$ near the Fermi momentum $k_F$ and about the $k$-dependence of the two-particle potential $g_{2,4}(k)$ -- the low-energy limit of interacting fermions in one dimension is described by the Tomonaga-Luttinger (TL) model. In absence of the spin degree of freedom (on which we focus in the work), the TL Hamiltonian reads\cite{tomonaga,luttinger}
\begin{equation}\label{eq:hTL}\begin{split}
H_\tn{TL}  = \sum_{n>0} \bigg[ k_n\left(v_F^0+\frac{g_4(k_n)}{2\pi}\right) & \left(b_n^\dagger b_n^{\phantom{\dagger}}
+ b_{-n}^\dagger b_{-n}^{\phantom{\dagger}} \right)\\
 +  k_n\frac{g_2(k_n)}{2\pi} & \left( b_n^\dagger b_{-n}^\dagger + b_{-n}^{\phantom{\dagger}} b_{n}^{\phantom{\dagger}}\right)\bigg]~,
\end{split}\end{equation}
where $b_n$ are free bosons, $k_n=2\pi n/L$, and $L$ denotes the system size. The low-energy equilibrium physics of $H_\tn{TL}$ is determined by two scalars, the LL parameter $K$ as well as the renormalized Fermi velocity $v_F$, which depend on $g_{2,4}(k=0)$; introducing canonical fields $\Pi$ and $\Theta$ yields
\begin{equation}
H_\tn{TL}^\tn{low energy} = \frac{v_F}{2} \int \left\{ K \left[\Pi(x)\right]^2+ K^{-1} \left[\partial_x\Theta(x)\right]^2 \right\}\,dx~.
\end{equation}
$H_\tn{TL}$ can be easily diagonalized, and after expressing fermionic annihilation operators in terms of the bosons $b_n$ one can derive analytical expressions for fermionic correlation functions at low energies. In a second step one needs to investigate about the validity of the initial assumptions that allowed to mathematically map interacting fermions to $H_\tn{TL}$. This is usually done by adding terms to the TL Hamiltonian which (partially) account for band curvature or physical choices of $g_{2,4}(k)$. One can then employ renormalization group (RG) arguments to show that for a very general class of models those terms are RG irrelevant and that the low-energy physics is indeed described by a TL fixed point theory with renormalized $K$ and $v_F$.\cite{haldane,RGmomentum,RGband,shankar}

Once a given Hamiltonian (one that might actually be motivated by an experimental setup and thus features a nonlinear dispersion, a finite bandwidth, and a physical two-particle interaction) was shown to fall into the LL universality class, the behavior of thermodynamic observables or correlation functions at low energies (or long length scales) is \textit{in principle} known; it is universally determined by the expressions derived within the TL model if the corresponding values for $K$ and $v_F$ are plugged in.\cite{haldane} However, there are only few quantitative verifications of Luttinger liquid physics \textit{directly for a microscopic theory}, i.e.~without resorting to the RG analysis within the TL fixed point model.\cite{volker} This particularly holds on the level of correlation functions -- the power law singularity of the momentum distribution $n(k)$, which is one of the key hallmarks of the LL phenomenology, has not been observed convincingly.\cite{dmrgnk,noacknk,ejimank,manmana} But even after a microscopic model was unambiguously shown to exhibit LL physics at low energies, other important questions remain: (i) What is the energy (or temperature) scale below which LL behavior manifests? What is the physics away from the low-energy (zero-temperature) limit? (ii) How can $K$ and $v_F$, which are non-universal functions of the model parameters, be obtained? (iii) How can the phase diagram be established? What happens to LL power laws if the system is driven out of the LL phase?

It is the first goal of this paper to address the above issues -- most importantly to unambiguously demonstrate the power law singularity of $n(k)$ -- for a lattice model of spinless fermions featuring nearest- as well as next-nearest neighbor interactions:
\begin{equation}\label{eq:h}\begin{split}
H = \sum_i \bigg[&\left( -\frac{1}{2}c_i^\dagger c_{i+1}^{\phantom{\dagger}} + \tn{H.c.}\right) \\
& + \Delta_{\phantom{2}} \left(c_i^\dagger c_i^{\phantom{\dagger}} -\frac{1}{2}\right) \left(c_{i+1}^\dagger c_{i+1}^{\phantom{\dagger}} -\frac{1}{2}\right)\\
& + \Delta_2 \left(c_i^\dagger c_i^{\phantom{\dagger}} -\frac{1}{2}\right) \left(c_{i+2}^\dagger c_{i+2}^{\phantom{\dagger}} -\frac{1}{2}\right) \bigg]~.
\end{split}\end{equation}
The ground state phase diagram for $\Delta_2=0$ can be computed analytically via Bethe ansatz.\cite{bethe1,bethe2} The half-filled system is a Luttinger liquid if $|\Delta|<1$. A gap opens for large interactions as charge density wave order ($\Delta>1$) or phase separation ($\Delta<-1$) develops. Away from half filling, the system is a LL for any repulsive $\Delta>0$. In presence of next-nearest neighbor interactions, no exact solution is known; it was shown numerically that the LL phase remains stable for small $\Delta_2$.\cite{nn1,nn2,nn3,nn4} One generally needs to resort to numerics (or approximate analytics) if one wants to calculate correlation functions which (even for $\Delta_2=0$) are hard to obtain by Bethe ansatz.\cite{bethespec,bethespec2} The density matrix renormalization group\cite{dmrgrev} (DMRG) for \textit{finite} systems was employed to investigate the fermionic momentum distribution,\cite{dmrgnk,noacknk,ejimank,manmana} the equal-time density response function\cite{schmitteckert,dmrgnk,noack,ejimapara,giamarchipara} as well as the density of states at zero temperature.\cite{dmrgdos1,dmrgdos2,dmrgdos3,jeckelmann} The results are consistent with the prediction from Luttinger liquid theory. The inverse system size, however, serves as an infrared cutoff, and despite substantial effort -- systems of up to $L=3200$ sites were studied in Ref.~\onlinecite{jeckelmann} -- very low energy (or long length) scales could not be accessed. To the best of our knowledge, fundamental LL power laws have not been observed unambiguously over several orders of magnitude for any microscopic model. 

It is the second goal of this paper to propose the use of an \textit{infinite}-system DMRG algorithm\cite{dmrgrev,tebd,selke} if one aims at studying Luttinger liquid physics at zero or finite temperature. Finite-size effects are absent by construction, and numerics are essentially controlled by one parameter $\chi$ (the dimension of the DMRG block Hilbert spaces) only; as a side effect, a calculation for an infinite system is computationally cheaper. We compute the single-particle Green function $G(x)$ and its Fourier transform, the momentum distribution function $n(k)$. At $T=0$, the anomalous decay of $G(x)\sim 1/x^{1+\alpha}$, or equivalently the power law singularity of $n(k)$ near the Fermi momentum $k_F$, $| n(k) -0.5 | \sim \left|k-k_F\right|^\alpha$, can be observed over several orders of magnitude; for $\Delta_2=0$, the exponent $\alpha$ is in perfect agreement with the TL formula\cite{voit,giamarchi} if $K$ from Bethe ansatz\cite{bethe1,bethe2} is plugged in. At finite temperatures, we show that $n(k)$ fulfills a simple scaling relation. We demonstrate that the energy scale on which the power law in $n(k)$ manifests vanishes when large $\Delta$ or $\Delta_2$ drive the half-filled system into a gapped phase; the behavior of $G(x)$ and $n(k)$ in the latter is discussed. We moreover compute the equal-time density response function $C_{NN}(x\tn{ or }k)$ at $T=0$ as well as the specific heat $c_V(T)$ and susceptibility $\xi(T)$; from these quantities, the LL parameters $K$ and $v_F$ can be obtained. We show how the different gapped phases\cite{nn4} and the associated type of order can be detected from the asymptotic behavior of correlation functions.

\begin{figure*}[t]
\includegraphics[width=0.46\linewidth,clip]{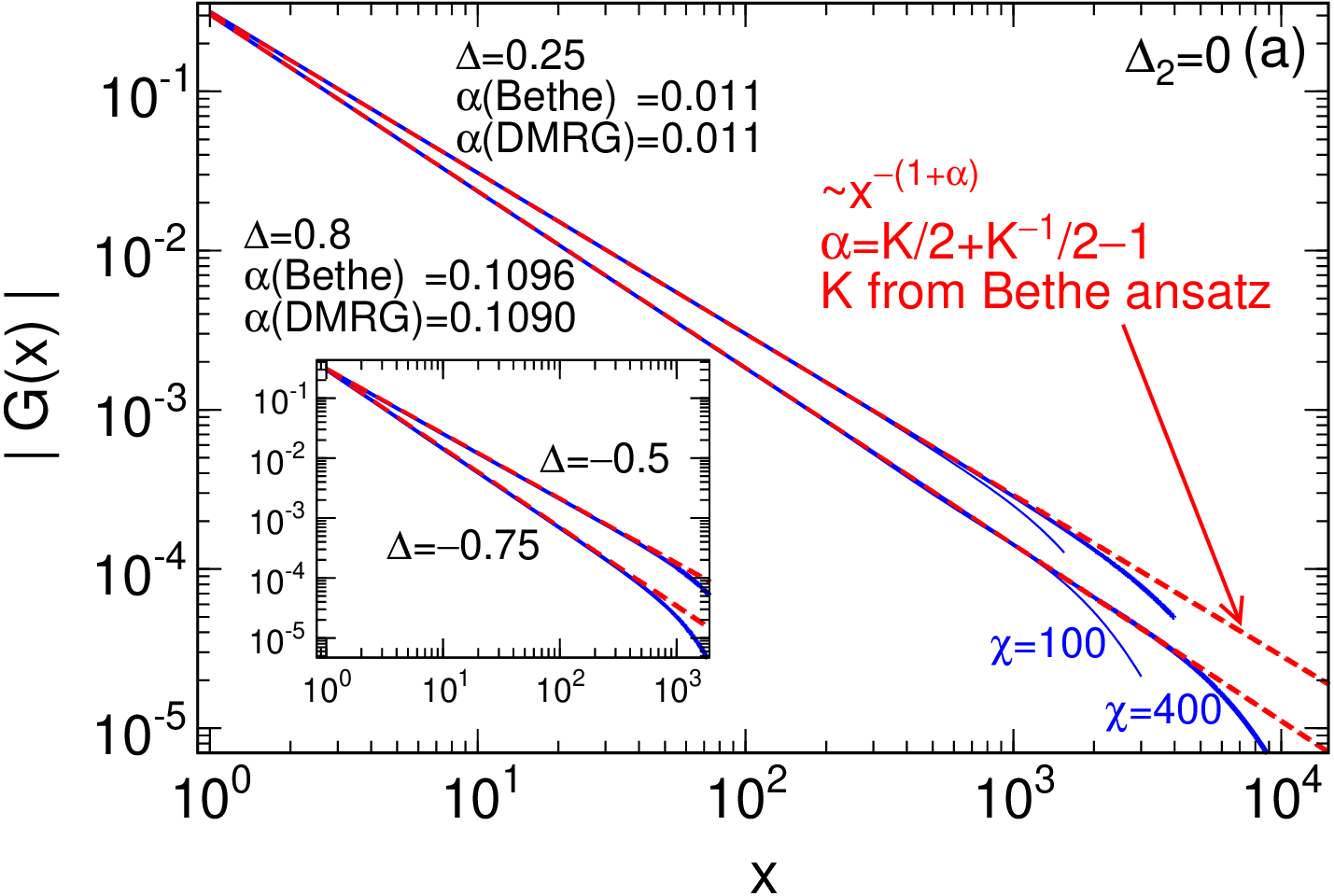}\hspace*{0.04\linewidth}
\includegraphics[width=0.46\linewidth,clip]{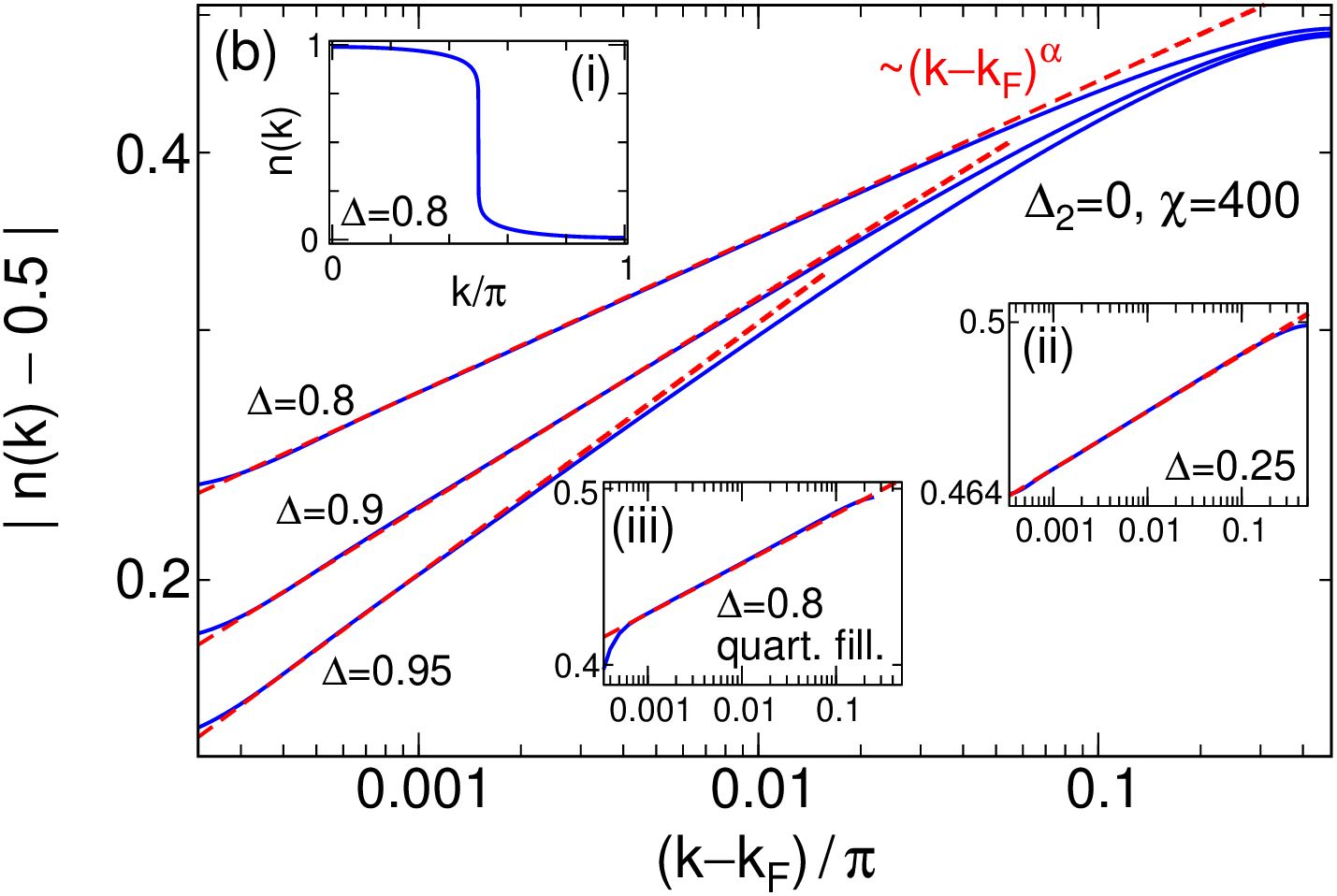}
\caption{(Color online) Zero-temperature single-particle Green function $G(x)$ and its Fourier transform, the fermionic momentum distribution function $n(k)$, for a half-filled [quarter-filled in (b,iii)] lattice of spinless fermions exhibiting a nearest neighbor interaction $\Delta$. Solid lines show DMRG data, dashed lines the asymptotic power laws $G(x)\sim 1/x^{1+\alpha(K)}$ and $|n(k)-0.5|\sim|k-k_F|^{\alpha(K)}$ expected for a Luttinger liquid. The LL parameter $K$ is taken from Bethe ansatz.\cite{bethe1,bethe2} We use an infinite-system DMRG algorithm where the `block Hilbert space dimension' $\chi$ is the only numerical controll parameter.\cite{commentnum} The LL power laws can be observed over several orders of magnitude. As the system is driven towards a gapped phase for $\Delta\to1$, the momentum scale on which they manifest vanishes. }
\label{fig:gf}
\end{figure*}

\section{Infinite-system DMRG}
\label{sec:model}

We study the equilibrium properties of Eq.~(\ref{eq:h}) at zero and finite temperature using a density matrix renormalization group algorithm\cite{white1,white2} for infinite systems.\cite{tebd,dmrgrev} Any translationally-invariant state can be expressed in terms of a matrix product state,\cite{mps}
\begin{equation}\label{eq:mps}\begin{split}
|\Psi\rangle = \sum_{\vec\sigma} \ldots & \left( A^{[1]{\sigma_{j+1\phantom{+N}}}} A^{[2]{\sigma_{j+2\phantom{+N}}}}\ldots A^{[N]{\sigma_{j+N\phantom{2}}}} \right) \\
\times & \left( A^{[1]{\sigma_{j+1+N}}} A^{[2]{\sigma_{j+2+N}}}\ldots A^{[N]{\sigma_{j+2N}}} \right)\ldots\left|\vec\sigma\right\rangle\,,
\end{split}\end{equation}
where $|\vec\sigma\rangle$ denotes a local product basis, characterized e.g. by occupation numbers or a spin-$1/2$ degree of freedom after applying a Jordan-Wigner transformation. The matrices $A^{[1\ldots N]\sigma_j}$ are associated with a unit cell of size $N$ (see below). At zero temperature, we determine the ground state $|\Psi_0\rangle$ by applying an imaginary time evolution $\exp(-\tau H)$ to a random initial state until the energy has converged to ten relevant digits. For $T>0$, one first needs to purify the thermal density matrix $\rho_T$ by introducing an auxiliary Hilbert space $Q$: $\rho_T = \tn{Tr\,}_{Q} |\Psi_T\rangle\langle\Psi_T|$. This is analytically possible only at $T=\infty$ where
\begin{equation}
\rho_\infty = \bigotimes_j \rho_{j,\infty}~,~~ \rho_{j,\infty} = \frac{1}{2}\sum_{\sigma_j}|\sigma_j\rangle\langle\sigma_j|~.
\end{equation}
However, $|\Psi_T\rangle$ can be obtained from $|\Psi_\infty\rangle$ by applying an imaginary time evolution, $|\Psi_T\rangle=e^{-H/(2T)}|\Psi_\infty\rangle$. \cite{dmrgT,barthel} Any static correlation function (on which we focus in this work) at zero or finite temperature can thus be \textit{exactly} expressed as
\begin{equation}\label{eq:gfab}
G^{AB}(j_1,j_2) =\langle A_{j_1}B_{j_2}\rangle_T =  \frac{ \langle \Psi_T| A_{j_1} B_{j_2} |\Psi_T\rangle}{ \sqrt{\langle\Psi_T|\Psi_T\rangle}}~. 
\end{equation}                                                                                                                                                                                                                                                                                                                                                                                                                                                                                                                                                                                                         
For an orthogonal matrix product state,\cite{dmrgrev} calculating the thermodynamic limit expectation value $\langle \Psi_T| A_{j_1} B_{j_2} |\Psi_T\rangle$ involves only a \textit{finite} number of sites $\sigma_j\in[j_1,j_2]$.

In order to implement the above procedure, we employ a standard time-dependent DMRG algorithm.\cite{tdmrg1,tdmrg2,tdmrg3,tdmrg4} After factorizing the evolution operators $\exp(-c\tau H)$ using a second order Trotter decomposition, they can be successively applied to Eq.~(\ref{eq:mps}). At each step $\Delta\tau$, singular value decompositions are carried out to update (some of the) matrices $A^{[1\ldots N]\sigma_j}$ within the unit cell. Since strict translational symmetry is broken by the Trotter decomposition,\cite{commentgap,commentunit} the latter is of size $N=2$ for $\Delta_2=0$ and $N=6$ otherwise.\cite{commentnum} At zero temperature, the matrix dimension $\chi$ is fixed and constitutes the numerical control parameter.\cite{commentnum} For $T>0$, $\chi$ is dynamically increased during the imaginary time evolution of $|\Psi_T\rangle$ such that at each step the sum of all squared discarded singular values is kept below a threshold value $\epsilon$; the DMRG approximation at finite temperature thus becomes exact in the limit $\epsilon\to0$ (and $\Delta\tau\to0$).

\section{Momentum distribution function}
\label{sec:gf}

The single-particle Green function and its Fourier transform (the fermionic momentum distribution function) are defined as\cite{commentnum}
\begin{equation}\label{eq:gf}
G(x) = \frac{1}{N}\sum_{y=1}^N\langle c^\dagger_{y+x}c^{\phantom{\dagger}}_{y}\rangle_T~,~~~n(k) = \sum_{x} e^{-ikx} G(x)~.
\end{equation}
Within the exactly-solvable Tomonaga-Luttinger model of Eq.~(\ref{eq:hTL}), $G(x)$ and $n(k)$ feature the following asymptotic zero-temperature power laws:\cite{voit,giamarchi,kurtrev1,kurtrev2}
\begin{equation}\label{eq:gfpl}\begin{split}
G(x\gg1) & \sim \frac{1}{x^{1+\alpha}}\,,~
 |n(k\approx k_F) -0.5 | \sim \left|k-k_F\right|^\alpha\,,
\end{split}\end{equation}
for parameters where $\alpha<1$ on which we focus in this work [for $\alpha>1$ a linear term dominates in $n(k)$; singularities appear in derivatives]. The anomalous dimension $\alpha$ is a function of the LL parameter $K$ only:
\begin{equation}\label{eq:alpha}
\alpha(K) = \frac{K}{2}+\frac{1}{2K}-1~.
\end{equation}
Exploiting universality, one expects that Eq.~(\ref{eq:gfpl}) also governs the long-distance (or low-energy) behavior of our spinless lattice fermions. However, $G(x)$ is hard to obtain by Bethe ansatz, and previous DMRG approaches\cite{dmrgnk,noacknk,ejimank,manmana} employed finite systems of $L=O(100)$ sites where potential power laws were cut off by the inverse system size $1/L$. To the best of our knowledge, nobody has succeeded in quantitatively demonstrating Eq.~(\ref{eq:gfpl}) -- which is a hallmark of LL physics -- for a microscopic model.

\begin{figure}[b]
\includegraphics[width=0.95\linewidth,clip]{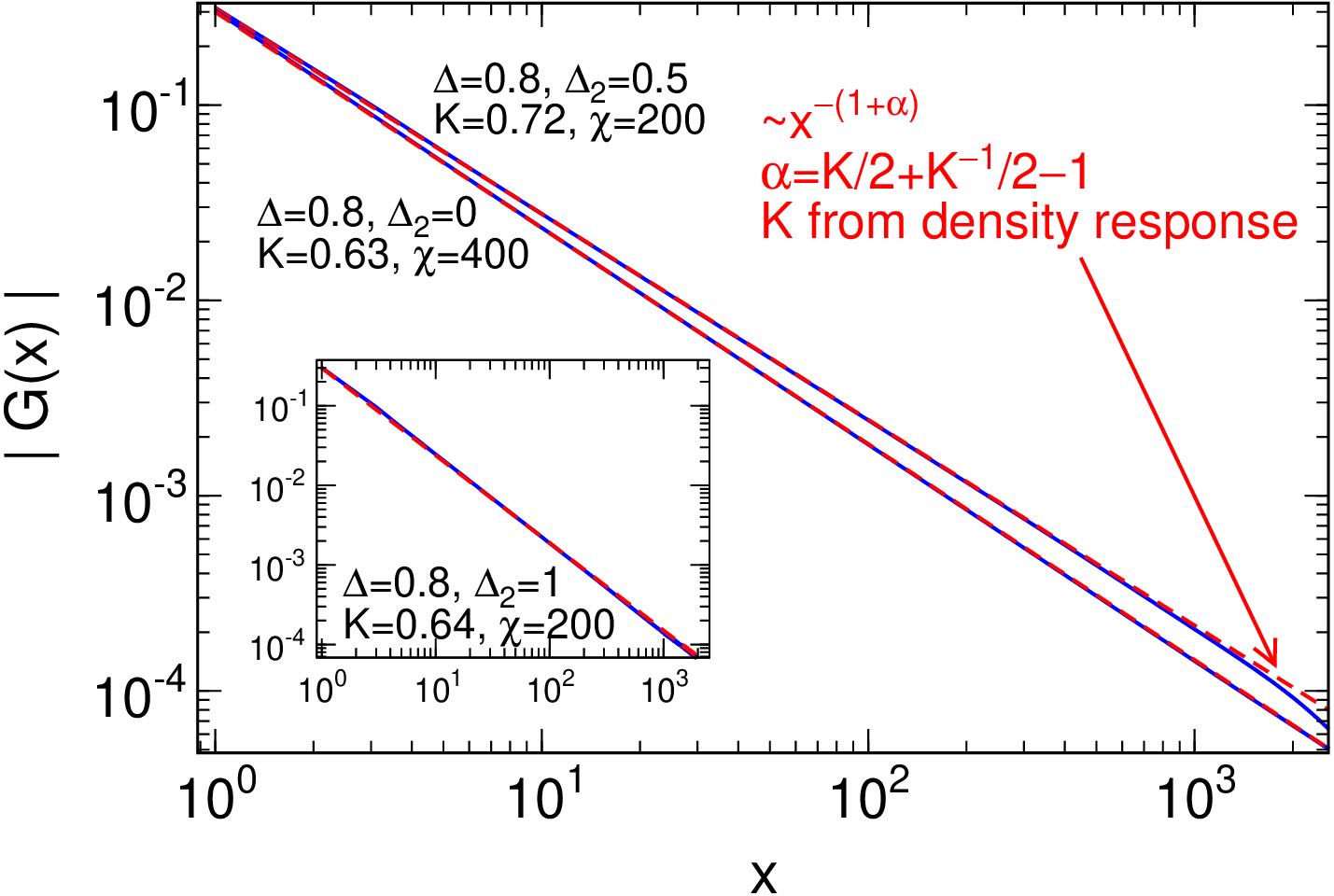}
\caption{(Color online) The same as in Figure \ref{fig:gf} but in presence of next-nearest neighbor interactions $\Delta_2$ which render the system nonintegrable. Dashed lines show the power law $\sim 1/x^{1+\alpha(K)}$ with the LL parameter $K$ determined from the density response function.  }
\label{fig:gfnn}
\end{figure}

\begin{figure*}[t]
\includegraphics[width=0.46\linewidth,clip]{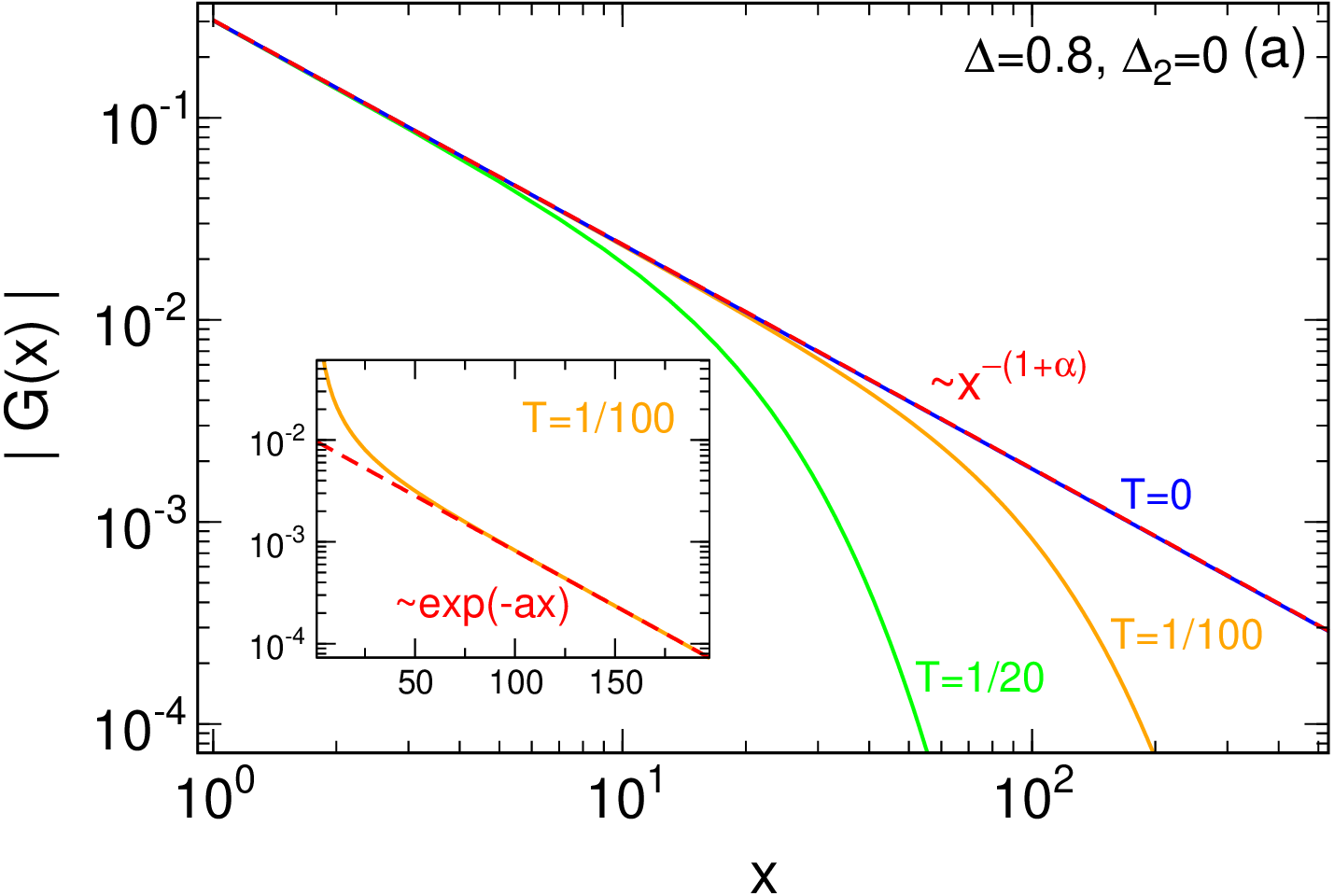}\hspace*{0.04\linewidth}
\includegraphics[width=0.46\linewidth,clip]{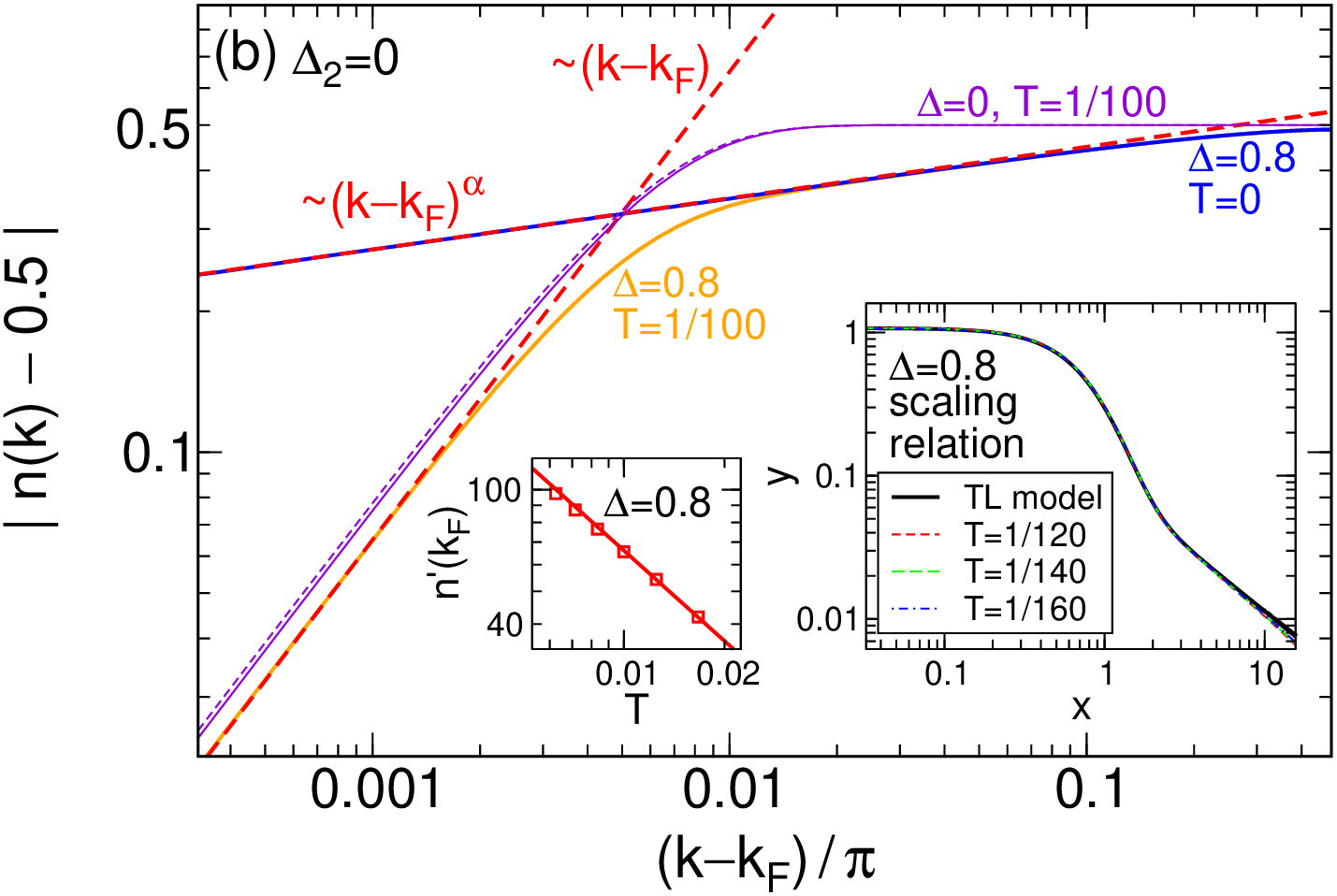}
\caption{(Color online) Single-particle Green function and momentum distribution at finite temperature $T$. The LL power laws are cut off at long distances (small momenta) and replaced by an exponential (linear) decay. The infinite-system DMRG algorithm is essentially controlled by the discarded weight $\epsilon=10^{-11}$ during the imaginary time evolution of the density matrix from $T=\infty$ down to the physical temperature. In (b), we compare our data (thin solid line) with the exact result (thin dashed line) for $\Delta=0$ and $T=1/100$ to demonstrate its accuracy. Left Inset to (b): Derivative $dn(k)/dk|_{k_F}$ as a function of temperature. The solid line shows the expected TL power law $\sim T^{\alpha-1}$. Right Inset to (b): Rescaled variables $y=T^{1-\alpha} dn(k)/d(\pi k)|_{k_F}$, $x=v_F(k-k_F)/(\pi T)$. Curves at different $T$ collapse and are consistent with a scaling relation obtained for the TL model. }
\label{fig:gfT}
\end{figure*}

Within our framework, $G(x)$ is directly accessible via Eqs.~(\ref{eq:gfab}) and (\ref{eq:gf}); $n(k)$ can be calculated by carrying out the Fourier sum in Eq.~(\ref{eq:gf}) until it has converged on the scale of the corresponding plot [typically, $O(10000)$ terms are necessary]. This allows us to verify the power laws of Eq.~(\ref{eq:gfpl}) and to investigate on which energy or temperature scale they manifest. We establish a scaling relation for $n(k,T)$ and provide indications that the latter is universal in the Luttinger liquid sense. 

\subsection{Zero temperature}

We show the zero-temperature single-particle Green function $G(x)$ as well as its Fourier transform $n(k)$ in Figures \ref{fig:gf} and \ref{fig:gfnn}. They feature asymptotic power laws over several orders of magnitude for any interaction strength $\Delta, \Delta_2$ and filling where the system is a Luttinger liquid [the filling is always one half except for Figure \ref{fig:gf}(b,iii) where it is one quarter]. For $\Delta_2=0$, the corresponding exponent is in perfect agreement with Eq.~(\ref{eq:alpha}) if $K$ from Bethe ansatz, which for half filling is given by
\begin{equation}\label{eq:K}
K = \frac{\pi}{2\arccos(-\Delta)}~,
\end{equation}
is plugged in (dashed lines in Figure \ref{fig:gf}). More quantitatively, a power law fit of $n(k)$ at $\Delta=0.8$ [at $\Delta=0.25$] yields an anomalous dimension $\alpha=0.109$ [$\alpha=0.011$] as compared to the exact result $\alpha(K_\tn{Bethe})=0.1096$ [$\alpha(K_\tn{Bethe})=0.011$]. In presence of next-nearest neighbor interactions, the model is nonintegrable and $K$ is not known exactly. In Section \ref{sec:llparam} we will demonstrate that it can be extracted accurately from the density response function at small momenta; if we determine $K$ for the parameters of Figure \ref{fig:gfnn} the corresponding exponent $\alpha(K)$ is consistent (dashed lines in Figure \ref{fig:gfnn}).

For small interactions, the power law decay of $n(k)$ is cut off in the ultraviolet only by the bandwidth [see Figure \ref{fig:gf}(b,ii)]. At half filling, large interactions drive the half-filled system into a gapped phase: $2k_F$ charge density wave order develops for $\Delta>1$, $\Delta_2=0$; if $\Delta_2>0$ is increased at any fixed $0\leq\Delta\leq1$, the system first enters a bond-ordered and eventually a $k_F$ charge density wave phase (see Refs.~\onlinecite{nn2,nn4} as well as Section \ref{sec:phase}). As one approaches the phase boundary, the scale on which the power law in $n(k)$ manifests vanishes. This is illustrated for $\Delta_2=0$ in Figure \ref{fig:gf}(b); note that even for $\Delta=0.95$ the power law is still observable over two orders of magnitude in our framework. In real space, the length scale above which $G(x)$ features a power law becomes larger for $\Delta$ close to $\Delta=1$. The overall behavior is similar if $\Delta_2$ is increased at fixed $\Delta$. The gapped phase itself will be discussed in Section \ref{sec:phase}.

\begin{figure}[t]
\includegraphics[width=0.95\linewidth,clip]{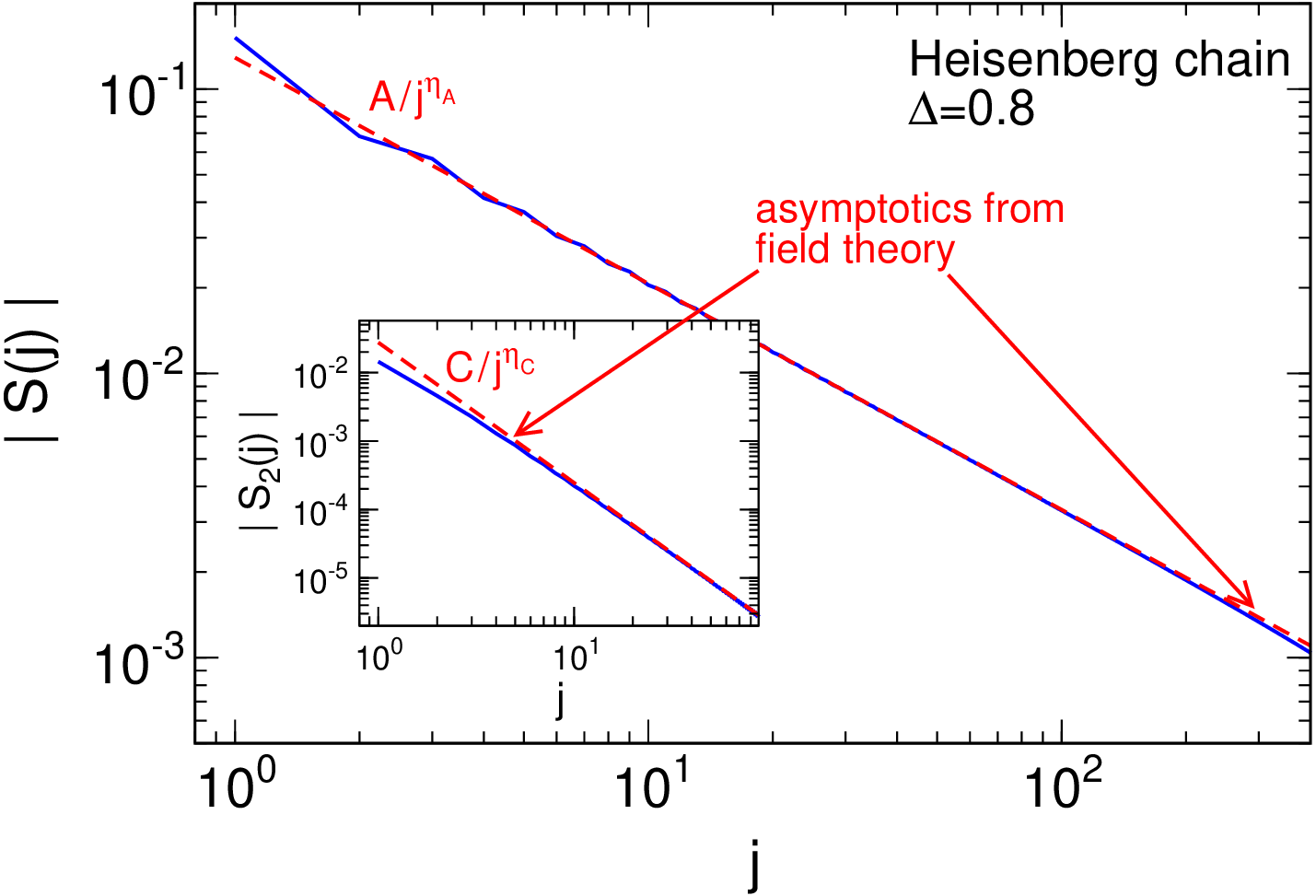}
\caption{(Color online) Zero-temperature spin correlation function $S(j)$ defined in Eq.~(\ref{eq:sj}) of a spin-$1/2$ Heisenberg chain with $z$-anisotropy $\Delta$. Solid lines shows DMRG data, dashed lines a field theory result of Ref.~\onlinecite{ft}. The latter yields both the exponents and coefficients of an asymptotic expansion at large distances. Inset: $S_2(j)=[\delta S(j)-\delta S(J+1)]/2$ with $\delta S(j) = S(j) - S_\tn{ft}^1(j)$ and $S_\tn{ft}^1(j)=A/j^\eta$ being the leading term within field theory. The difference $S_2(j)$ is governed by the leading alternating term.}
\label{fig:chain}
\end{figure}

Within our zero-temperature algorithm, the matrices in Eq.~(\ref{eq:mps}) which describe the ground state are determined through an imaginary time evolution of a random initial state.\cite{dmrgrev,tebd,commentnum} The key approximation results from the finite fixed matrix dimension $\chi$ in the state constructed by the algorithm (which leads to an effective correlation length that diverges as a power law in $\chi$).\cite{tebd2,tebd3} However, no finite-$\chi$ extrapolation is necessary for the results of this paper as reasonable matrix sizes already yield Luttinger liquid behavior over a broad range: Even moderate values $\chi=100-400$ are sufficient to observe the power laws in $G(x)$ or $n(k)$ over several orders of magnitude (see Figure \ref{fig:gf}). It should be emphasized again that $\chi$ is the only numerical control parameter and that finite size effects are by construction absent in our approach. From a physical perspective, the use of an infinite-system algorithm is motivated by the fact that in a finite system one would expect power laws to be cut off in the infrared by the inverse system size $1/L$. Indeed, if we pragmatically restrict the Fourier sums in Eq.~(\ref{eq:gf}) to $O(100)$ sites -- which is a typical value for a finite-system DMRG calculation\cite{dmrgnk,noacknk,ejimank,manmana} -- no power law behavior is visible; it is moreover instructive to directly compare our Figure \ref{fig:gf}(b) with Figure 3 of Ref.~\onlinecite{ejimank} which was obtained using $L=66$.

\begin{figure*}[t]
\includegraphics[width=0.46\linewidth,clip]{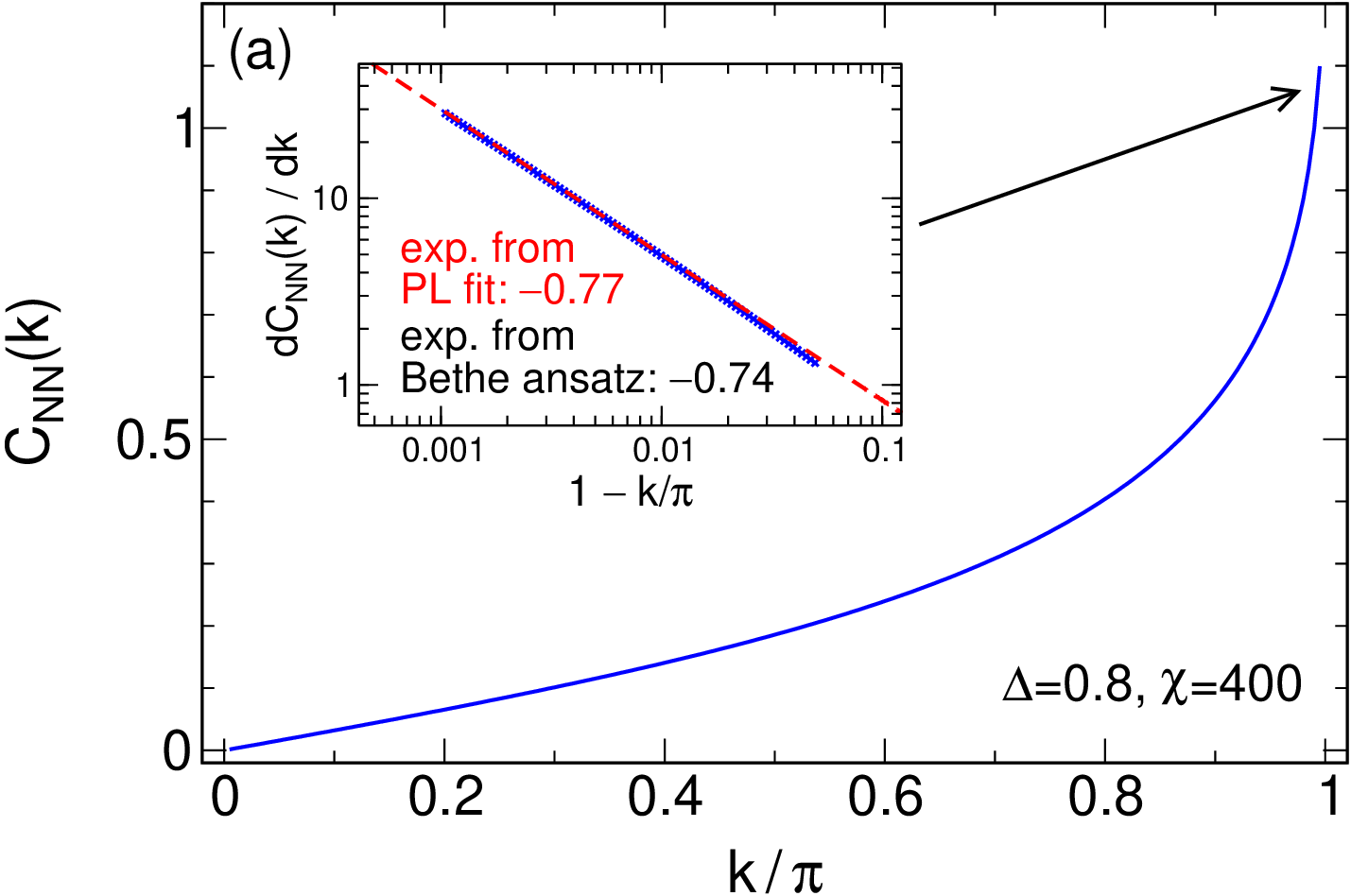}\hspace*{0.04\linewidth}
\includegraphics[width=0.46\linewidth,clip]{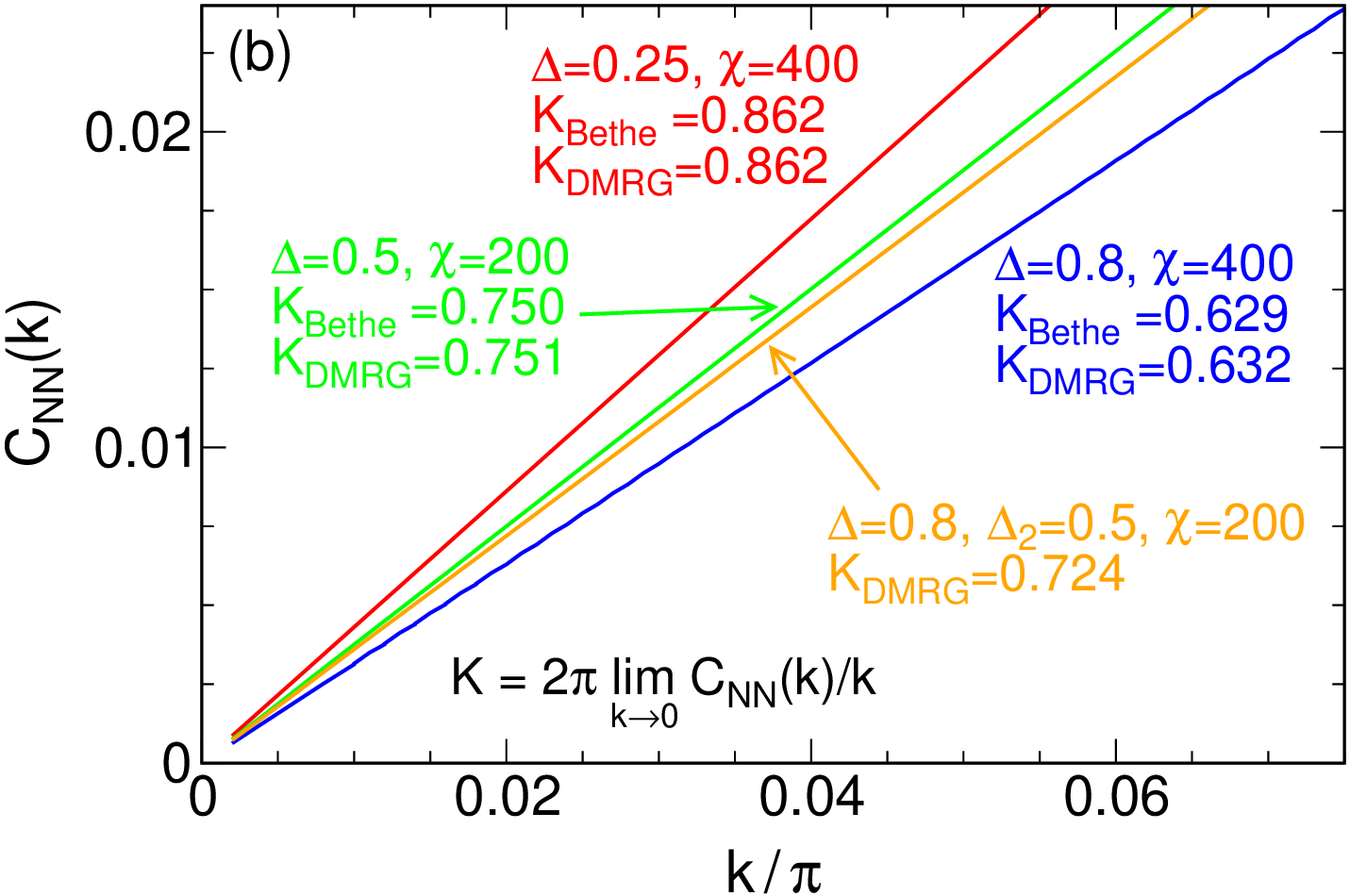}
\caption{(Color online) DMRG calculation of the zero-temperature instantaneous density response function $C_{NN}(x)$. Its Fourier transform $C_{NN}(k)$ exhibits a power law singularity $C_{NN}(k\approx 2k_F)\sim (2k_F-k)^{2K-1}$ for momenta near $2k_F$. For small $k$, $C_{NN}(k)$ is linear with a slope determined by the Luttinger liquid parameter $K$. The latter can thus be determined accurately through linear fits [see the comparison with Bethe ansatz results in (b)].}
\label{fig:density}
\end{figure*}

\subsection{Finite temperature}

We now turn to finite temperatures. This is interesting since (a) LL power laws will be cut off in the infrared by $T$; thus, one needs to study on which temperature scale they are still observable for a given microscopic model (e.g.~one that could be motivated by an experimental setup), (b) little is known about $n(k,T)$ if both $|k-k_F|\neq0$ and $T\neq0$; due to the absence of any low energy scale one might expect a scaling relation for $n(k,T)$, and (c) adressing those questions establishes the infinite-system DMRG algorithm as a convenient tool to study systems at \textit{nonzero temperature} directly in the thermodynamic limit.

Let us briefly recall some computational details: The single-particle Green function $G(x)$ (or any other correlation function) can be calculated staightforwardly via Eq.~(\ref{eq:gfab}), and the state $|\Psi_T\rangle$ which purifies the density matrix is obtained from the one at $T=\infty$ (where it is known) through an imaginary time evolution down to the physical temperature $T$.\cite{dmrgT,barthel} The computational effort is moderate since only $N=4$ (for $\Delta_2=0$; $N=12$ for $\Delta_2>0$) different matrices appear in Eq.~(\ref{eq:mps}). Numerics are controlled by the discarded weight $\epsilon$ during each Trotter step as well as by the step size $\Delta\tau$. We typically choose $\epsilon=10^{-11}$ and $\Delta\tau=0.01$. A comparison with the analytic result at $\Delta=\Delta_2=0$ and $T=1/100$ shows that this is sufficient to obtain accurate data [see Figure \ref{fig:gfT}(b)].

Our results are shown in Figure \ref{fig:gfT}. Power laws are cut off by the temperature; $G(x)$ decays exponentially at large lengths, and $n(k)$ is linear for momenta close to $k_F$. For $T\lesssim 1/100$ and $\Delta=0.8$, the LL power law in $n(k)$ is observable over about one order of magnitude in $|k-k_F|$. Due to the absence of any low energy scale except for $v_F|k-k_F|$ and $T$, one also expects power laws in temperature. Indeed, for the TL model the $k$-derivative satisfies\cite{scalingfunction} 
\begin{equation}
\frac{dn(k)}{dk}\bigg|_{k_F}\sim~ T^{\alpha-1} ~,
\end{equation}
which is consistent with our data for the lattice model [see the left inset to Figure \ref{fig:gfT}(b)]. For $|k-k_F|\neq0$ and $T\neq 0$ one can show that within the TL model $n(k,T)$ has a simple scaling form
\begin{equation}\label{eq:scaling}
\frac{dn(k)}{dk} \sim T^{\alpha-1} F\left[\frac{v_F(k-k_F)}{\pi T} \right]~.
\end{equation}
Eq.~(\ref{eq:scaling}) can be established (and the scaling function $F$ can be computed) by expressing $n(k,T)$ in terms of the local density of states which within the TL model was shown to have scaling form.\cite{scalingfunction} In the right inset to Figure \ref{fig:gfT}(b) we illustrate that our DMRG data for the lattice fermions at small $T$ and $|k-k_F|$ agrees with Eq.~(\ref{eq:scaling}) and with the form of $F$ (if in the latter both axes are scaled arbitrarily). This indicates that the low-energy behavior of $n(k,T)$ at finite temperature is universal for any model that falls into the Luttinger liquid universality class. 

\begin{figure}[b]
\includegraphics[width=0.95\linewidth,clip]{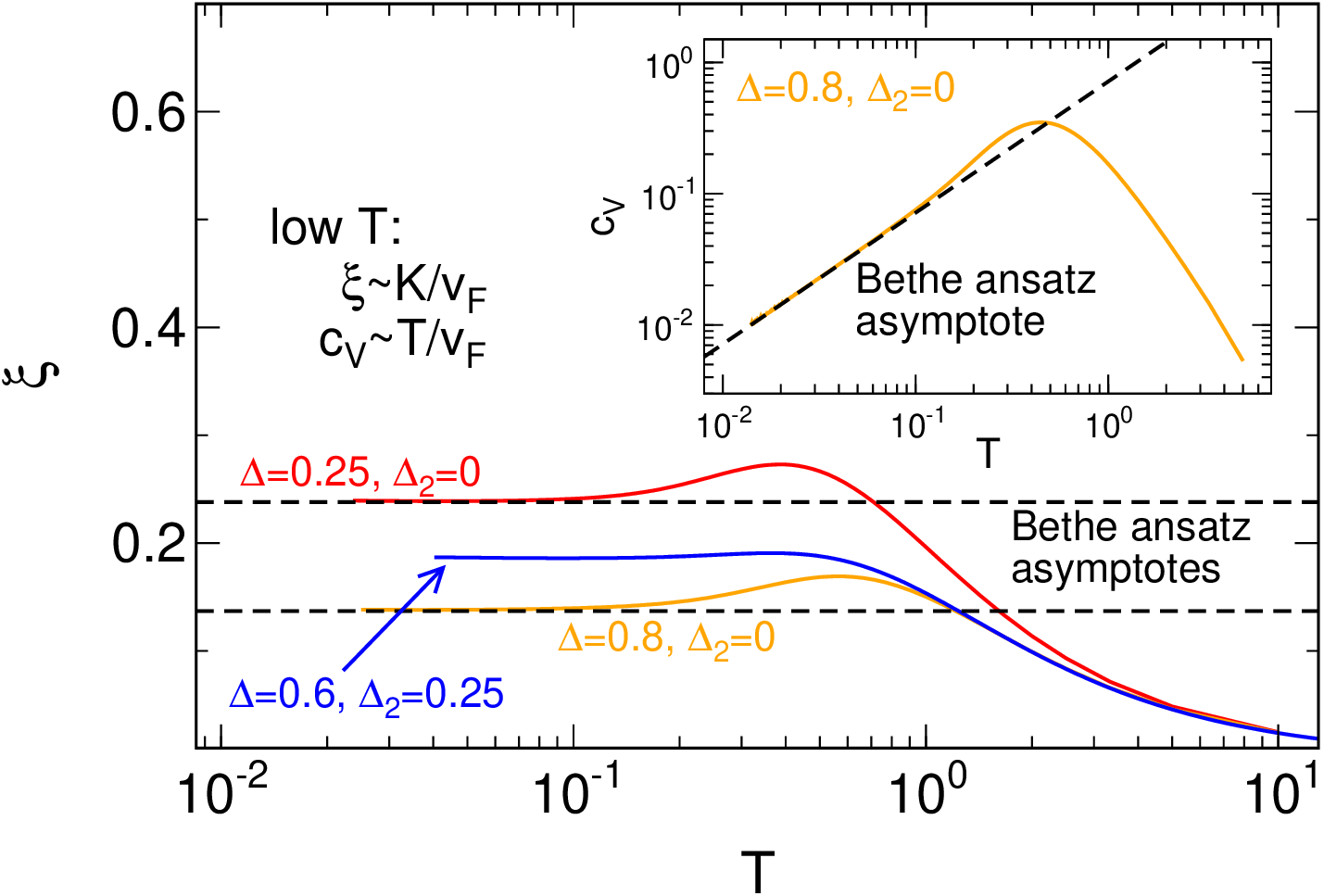}
\caption{(Color online) Thermodynamic quantities from infinite-system DMRG. The main panel shows the susceptibility defined in Eq.~(\ref{eq:chi}); the inset displays the specific heat $c_V(T)$. The low-temperature behavior is determined by the LL parameter $K$ and renormalized Fermi velocity $v_F$. At $\Delta_2=0$, the dashed lines show the asymptotic behavior predicted by Bethe ansatz.\cite{betheT} Our approach provides a way to extract $K$ and $v_F$ for nonintegrable models.}
\label{fig:td}
\end{figure}

\subsection{Spin chain}

For the integrable model, we have shown that $G(x)$ asymptotically decays as $\sim 1/x^{1+\alpha(K)}$ if the LL parameter $K$ from Bethe ansatz is plugged in. We will now investigate if one can make a reasonable statement about \textit{subleading} terms. To this end, we consider chain of spin-$1/2$ degrees of freedom $S^{x,y,z}$:
\begin{equation}
H_s = \sum_j \left(S^x_jS^x_{j+1} + S^y_jS^y_{j+1} + \Delta S^z_jS^z_{j+1} \right) ~,
\end{equation}
which can be mapped to Eq.~(\ref{eq:h}) through a Jordan-Wigner transformation
\begin{equation}\label{eq:jw}
S^+_j = S^x_j + iS^y_j = c_j^\dagger e^{i\pi\sum_{k=-\infty}^{j-1}c_j^\dagger c_j^{\phantom{\dagger}}}~.
\end{equation}
For the longitudinal correlation function\cite{commentunit}
\begin{equation}\label{eq:sj}
S(j)=\frac{1}{N}\sum_{k=1}^N\langle S^x_{k+j}S^x_k\rangle_T
\end{equation}
which is the spin equivalent of the single particle Green function of Eq.~(\ref{eq:gf}) -- an asymptotic expansion
\begin{equation}
S(j) \sim \frac{A}{j^{\eta_A}} + \frac{B}{j^{\eta_B}} + (-1)^j \frac{C}{j^{\eta_C}} + \ldots
\end{equation}
was obtained by field theory,\cite{ft} and the exponents as well as the \textit{prefactors} were computed. One might speculate that the string operators in Eq.~(\ref{eq:jw}) hinder a similar approach in the fermionic case.

In Figure \ref{fig:chain} we show that zero-temperature DMRG calculations for $S(j)$ at $\Delta=0.8$ compare nicely with the field theory result.\cite{dmrgft} The exponent of the leading term $\eta_A$ is much smaller than $\eta_{B,C}$ and thus dominates at large lengths. If we substract this term, $\delta S(j) = S(j) - A/j^{\eta_A}$, and compute the difference $S_2(j)=[\delta S(j)-\delta S(j+1)]/2$, the leading alternating term $(-1)^jC/\eta_C$ can clearly be observed. The subleading homogeneous term $B/j^{\eta_B}$ is consistent with the behavior of $S_2^+(j)=[\delta S(j)+\delta S(j+1)]/2$ but cannot be detected with similar accuracy (because any slight derivation from the substracted field theory exponent $\eta_A$ significantly modifies $S^+_2$ but not the difference $S_2$).

\section{Extraction of LL parameters}
\label{sec:llparam}

In this Section we discuss further static correlation functions as well as thermodynamic quantities. Our main focus will be to demonstrate how the Luttinger liquid parameter $K$ as well as the renormalized Fermi velocity $v_F$ -- which fully characterize the TL model and thus LL physics -- can be extracted.\cite{ejimapara,giamarchipara,lehurpara} For $\Delta_2=0$ we achieve excellent agreement with the exact Bethe ansatz result. Our algorithm thus provides an efficient way to determine $K$ and $v_F$ for a nonintegrable model.

\subsection{Density response}

The two-particle analog of $G(x)$ is the instantaneous density response function\cite{commentunit}
\begin{equation}\label{eq:cnn}\begin{split}
C_{NN}(x) & = \frac{1}{N}\sum_{y=1}^N\left(\langle c_{y+x}^\dagger c_{y+x}^{\phantom{\dagger}} c_{y}^\dagger c_{y}^{\phantom{\dagger}}\rangle_T
- \langle c_{y+x}^\dagger c_{y+x}^{\phantom{\dagger}}\rangle_T\langle c_y^\dagger c_y^{\phantom{\dagger}}\rangle_T\right)~, \\
C_{NN}(k) & = \sum_x e^{-ikx}C_{NN}(x)~.
\end{split}\end{equation}
Within the Tomonaga-Luttinger model, it exhibits the following characteristic zero-temperature behavior,\cite{voit,giamarchi}
\begin{equation}\label{eq:cnnplx}
C_{NN}(x) = \frac{K}{2\pi^2 x^2} + C\frac{\cos(2k_Fx)}{x^{2K}}+\ldots~,
\end{equation}
where $C$ is a nonuniversal constant. In momentum space this translates to
\begin{equation}\label{eq:cnnplk}
\frac{d C_{NN}(k)}{dk} =
\begin{cases}
K/(2\pi) & k=0 \\
\sim (2k_F-k)^{2K-2} & k\approx 2k_F ~.
\end{cases}
\end{equation}
DMRG data for the spinless lattice fermions\cite{dmrgnk,noack,ejimapara,giamarchipara} at half filling is shown in Figure \ref{fig:density}. We reproduce the power law singularity for momenta close to $2k_F$; a power law fit at $\Delta=0.8, \Delta_2=0$ yields an exponent of $-0.77$ which is in good agreement with the exact result $2K_\tn{Bethe}-2\approx-0.74$. The slope at small $k$ is determined by the LL parameter $K$ which can thus be determined accurately from simple \textit{linear} fits. Even for moderate values of $\chi=200$, $K$ agrees perfectly with the exact Bethe ansatz result Eq.~(\ref{eq:K}) for $\Delta_2=0$ [see the comparisons at $\Delta=0.25, 0.5, 0.8$ in Figure \ref{fig:density}(b)]. Our algorithm thus provides a rather efficient way to compute $K$ for nonintegrable models.

In contrast to the momentum distribution function, the density response at small momenta was calculated reliably before using DMRG for finite systems.\cite{ejimapara} It turned out that using $L\approx100-400$ sites is sufficient to extrapolate to the thermodynamic limit and to extract $K$, which seems reasonable since the first term in Eq.~(\ref{eq:cnnplx}) falls off rapidly. Depending on the problem at hand, our approach might be more efficient (calculations at $\Delta_2=0$ and $\chi=200$ take about an hour).

\subsection{Thermodynamic quantities}

We finally consider two thermodynamic quantities: the specific heat (per site)\cite{voit,giamarchi}
\begin{equation}
c_V(T) = \frac{d\langle H\rangle_T}{dT}\, \stackrel{T\to0}{\sim}\, \frac{T}{v_F}
\end{equation}
as well as the susceptibility (we use the terminology `susceptibility' following previous works on the Heisenberg chain)\cite{chi}
\begin{equation}\label{eq:chi}
\xi(T) = \frac{1}{T}\frac{1}{N}\sum_{k=1}^N\sum_j \left\langle S^z_{k+j}S^z_{k} \right\rangle_T
\stackrel{T\to0}{\sim} \frac{K}{v_F}~, 
\end{equation}
where $S^z_j=c_j^\dagger c_j-1/2$. Both the LL parameter $K$ and the renormalized Fermi velocity $v_F$ can be determined from the low-temperature behavior of $c_V/c_V^0$ and $\xi/\xi^0$, where the ratio with the noninteracting (known) values $c_V^0$ and $\xi^0$ is taken so that nonuniversal prefactors drop out. In Figure \ref{fig:td}, we show $c_V$ and $\xi$ in comparison with the Bethe ansatz asymptotes\cite{betheT} for $\Delta_2=0$. The low-energy regime can be reached easily, and the agreement with the exact result is quantitative. Thus, our algorithm at nonzero temperature provides an accurate means to determine $K$ and $v_F$ for nonintegrable models.\cite{giamarchipara}

\begin{figure}[b]
\includegraphics[width=0.95\linewidth,clip]{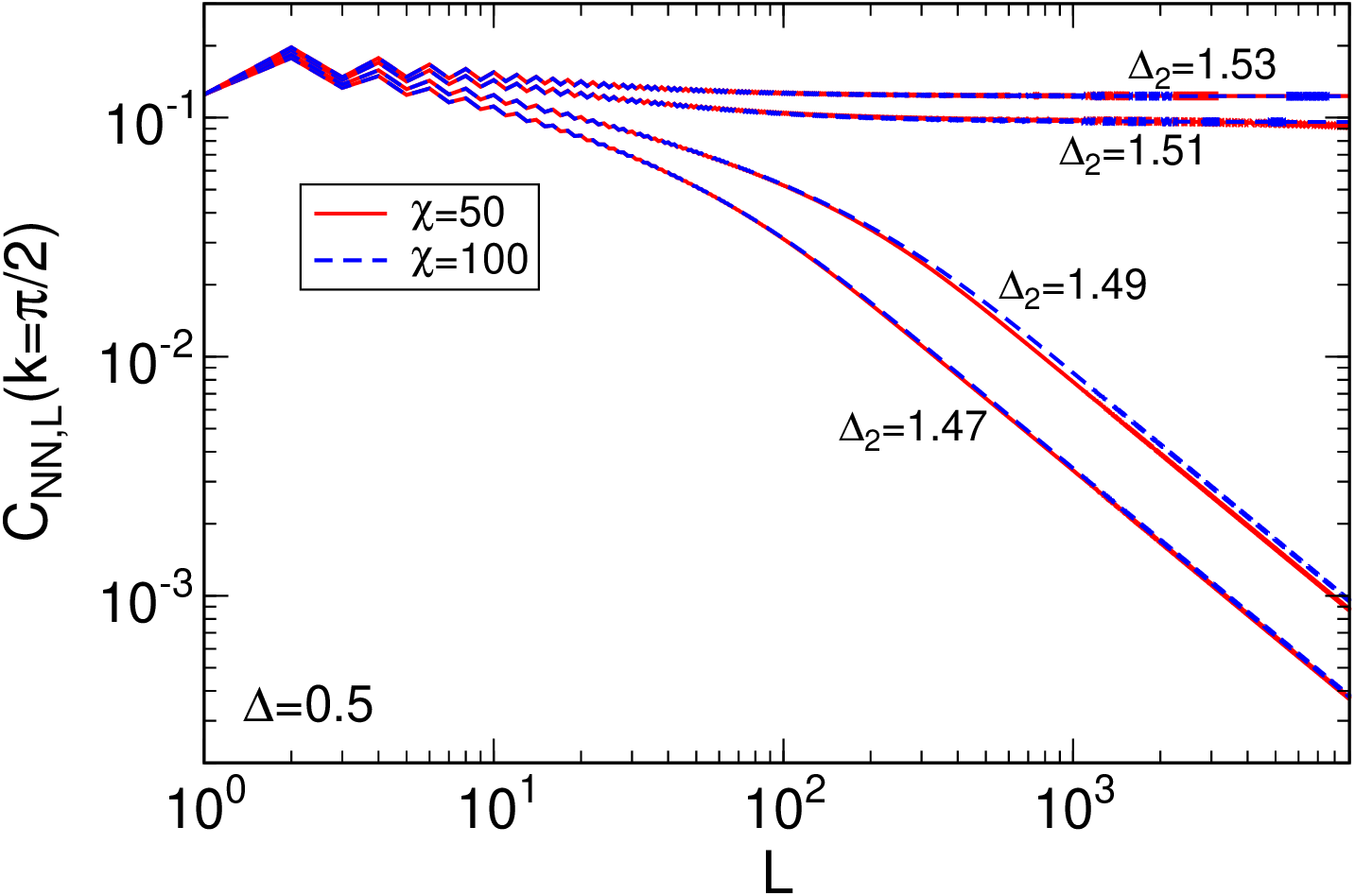}
\caption{(Color online) Density-density correlation function $C_{NN,L}(k) = L^{-1}\sum_{x=-L}^L e^{-ikx}C_{NN}(x)$. $C_{NN,L}(k_F=\pi/2)$ vanishes for $\L\to\infty$ in the bond-ordered phase but remains finite in the $k_F$ charge density wave phase. Moderate values of $\chi=100$ predict that the transition occurs at $\Delta_2=1.50\pm0.01$ in agreement with $\Delta_2=1.5$ obtained in Ref.~\onlinecite{nn4} (the transition from the LL to the bond-ordered phase takes place at $\Delta_2=1.25$).}
\label{fig:phase}
\end{figure}

\begin{figure*}[t]
\includegraphics[width=0.46\linewidth,clip]{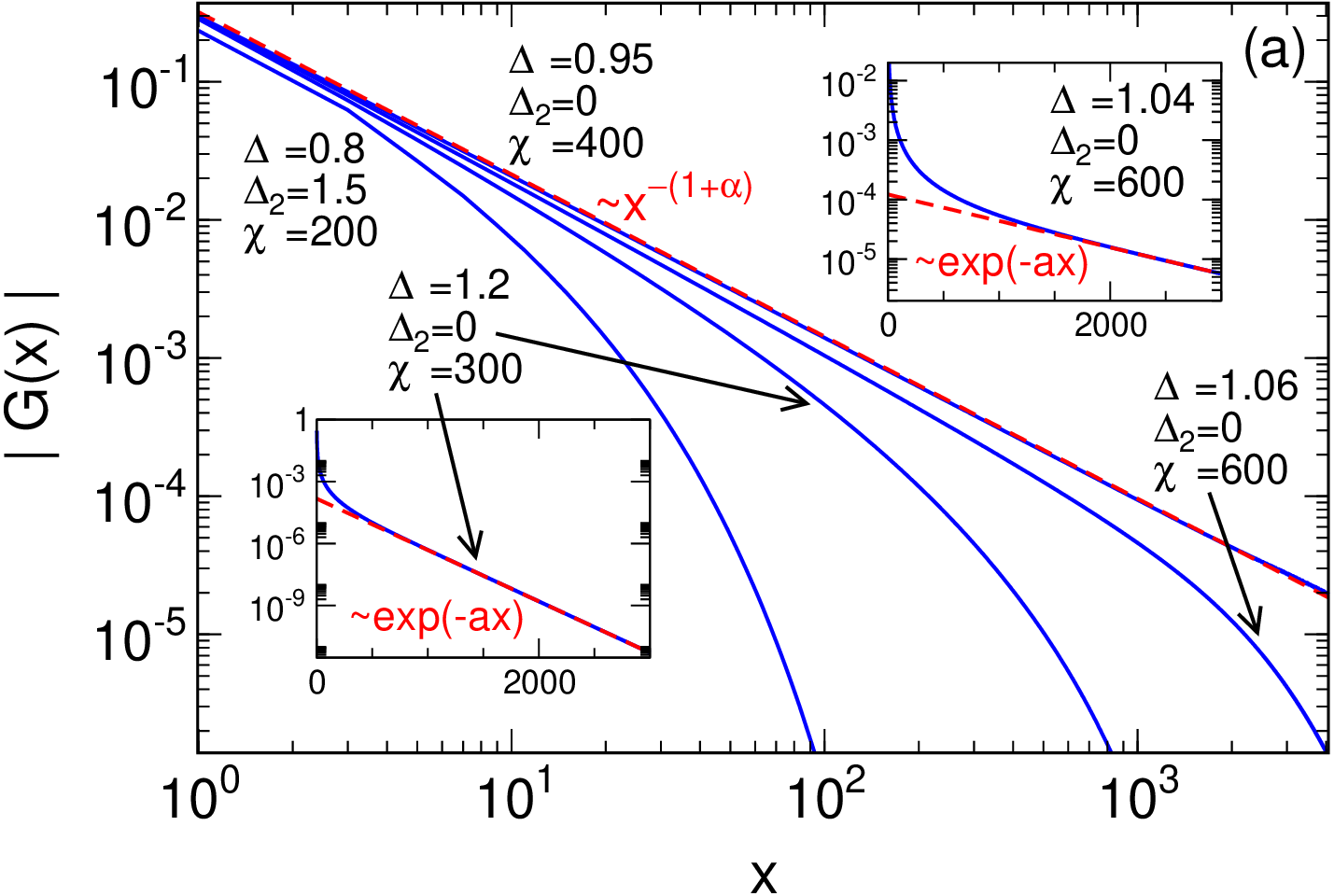}\hspace*{0.04\linewidth}
\includegraphics[width=0.46\linewidth,clip]{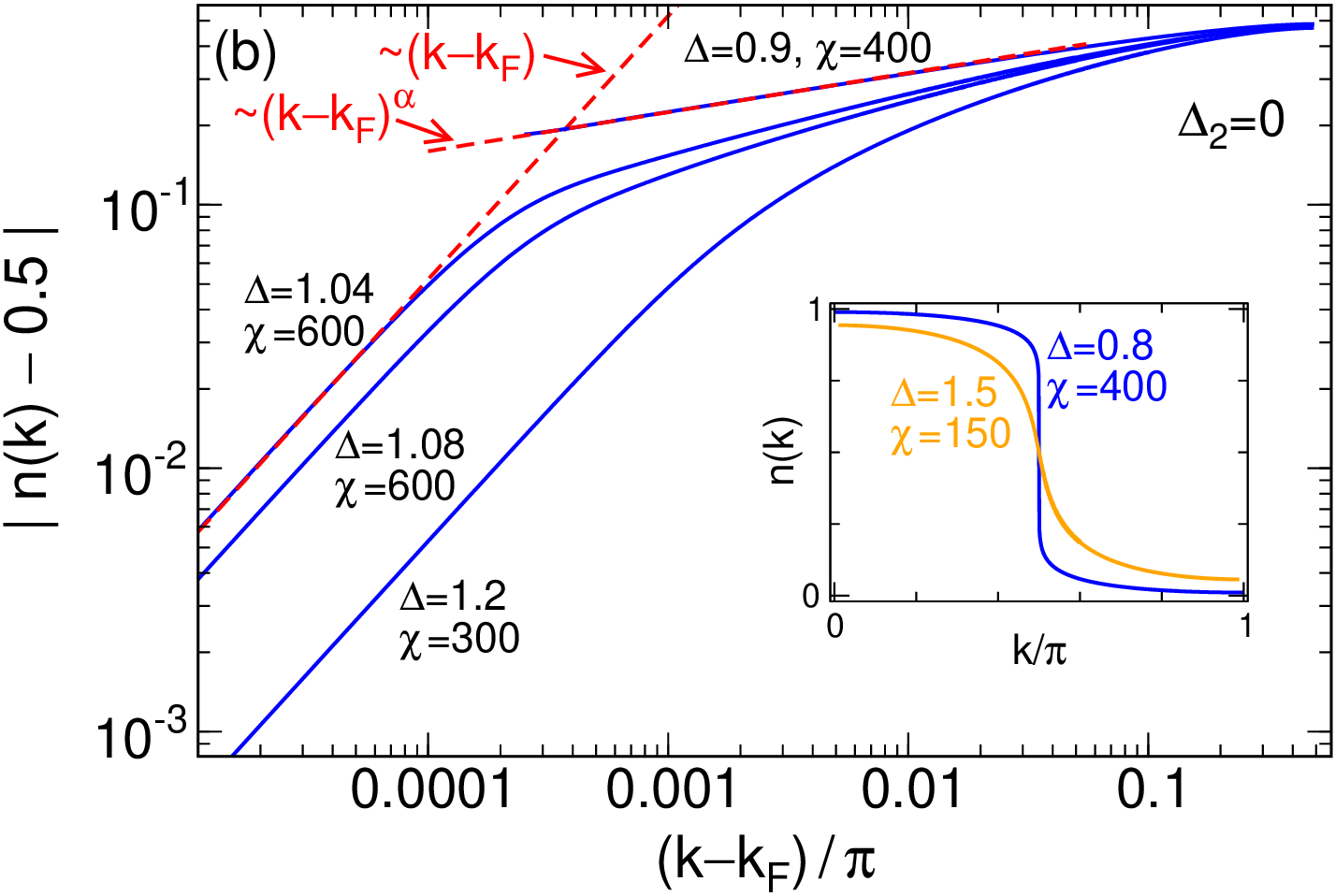}
\caption{(Color online) Single-particle Green function and momentum distribution for large interactions that drive the half-filled system into gapped phases (a $2k_F$ charge density wave phase for $\Delta>1$, $\Delta_2=0$ and a bond-ordered phase at $\Delta=0.8$, $\Delta_2=1.5$; see Ref.~\onlinecite{nn4} as well as Section \ref{sec:phase} for details). $G(x)$ decays exponentially, and $n(k)$ is linear for close to $k_F$. $\chi$ is chosen such that the results are converged on the scale of this plot.}
\label{fig:gfgap}
\end{figure*}

\section{Phase diagrams}
\label{sec:phase}

The phase diagram for our model\cite{nn1,nn3} at $\Delta_2>0$ was determined accurately in Refs.~\onlinecite{nn2} and \onlinecite{nn4} using finite-system DMRG. The Luttinger liquid phase remains stable for finite next-nearest neighbor interactions. As $\Delta_2$ increases, the system is first driven into a gapped bond-ordered phase and then into a $k_F$ charge density wave phase. We can reproduce these results by considering the asymptotic behavior of the corresponding correlation function.\cite{commentgap} This is exemplified in Figure \ref{fig:phase}: The density-density correlation
\begin{equation}
C_{NN,L}(k) = \frac{1}{L} \sum_{x=-L}^L e^{-ikx} C_{NN}(x)
\end{equation}
at $k=k_F$ vanishes for $L\to\infty$ in the bond-ordered phase but is finite in the $k_F$ charge density wave phase. Even moderate values of $\chi=100$ predict the transition to occur at $\Delta_2=1.50\pm0.01$ (for $\Delta=0.5$) in agreement with $\Delta_2=1.5$ obtained in Ref.~\onlinecite{nn4}. To detect the transition out of a LL phase it is reasonable to consider the single-particle Green function $G(x)$ in real space. Its LL power law behavior is replaced by an exponential decay in the $2k_F$ charge density wave (e.g., $\Delta\gg1$) or bond-ordered phase (see Figure \ref{fig:gfgap}). However, the effort to accurately detect the phase boundary of the Kosterlitz-Thouless transition (where a gap opens exponentially) is significant [$\chi=600$ is necessary to establish that $G(x)$ falls off exponentially at $\Delta=1.04$], and the approach of Ref.~\onlinecite{nn4} seems more efficient. The advantage of our method is that correlation functions -- and thus the type of order -- can be obtained easily (establishing a Figure analogous to our Figure \ref{fig:phase} is much harder within the finite-system DMRG approach of Ref.~\onlinecite{nn4}).

\section{Conclusion and outlook}

We studied zero- and finite-temperature Luttinger liquid physics of spinless lattice fermions exhibiting nearest and (nonintegrable) next-nearest neighbor interactions terms $\Delta$ and $\Delta_2$. Employing an infinite-system density matrix renormalization group algorithm (and carrying out Fourier sums up to $\sim10000$ sites), we unambiguously showed that the momentum distribution function $n(k)$ at $T=0$ exhibits the characteristic power law suppression near the Fermi momentum. For $\Delta_2=0$, the associated exponent is in perfect agreement with the one predicted for the Tomonaga-Luttinger model if the corresponding LL parameter $K$ is taken from Bethe ansatz. The scale on which this power law is visible vanishes when large interactions drive the half-filled system into a gapped phase. Finite temperatures cut off power laws in the infrared, and $n(k,T)$ obeys a scaling relation. We illustrated that the LL parameter $K$ and renormalized Fermi velocity $v_F$ are easily extracted from the instantaneous density response at small momenta, from the specific heat, or from the susceptibility. Phase transitions (and the associated types of order) can be detected from the asymptotics of correlation functions. Both is of importance for models which cannot be solved analytically. This overall indicates that employing infinite-system DMRG might be advantageous if one aims at investigating Luttinger liquid physics of (homogeneous) microscopic models at low energies or temperatures. As a next logical step it would be interesting to calculate \textit{time-dependent correlation functions}, and their Fourier transform, the local density of states.

\emph{Acknowledgments} --- We are indebted to Volker Meden and Kurt Sch\"onhammer for fruitful discussions and comments and acknowledge support by the Deutsche Forschungsgemeinschaft via KA3360-1/1 (C.K.) as well as by the AFOSR MURI on ``Control of Thermal and Electrical Transport'' (J.E.M.).

%%%%%%%%%%%%%%%%%%%%%%%%%%%%%%%%%%%%%%%%   Bibliography


\begin{thebibliography}{*}
\bibitem{commentgap}
A degenerate ground state (which occurs e.g.~for $\Delta>1$, $\Delta_2=0$) is signalled by the fact that the matrices in the unit cell are unequal at the end of the imaginary time evolution where the Trotter step $\Delta\tau\to0$. Our algorithm randomly picks one of those ground states; the other ones can be simply obtained by shifting the unit cell (or successively re-running the calculation). To ensure that this is correct one can compare the results with the limit $T\to0$ of a finite-temperature calculation (which by construction accounts for any state of the system). We have exemplary studied that the finite-$T$ single-particle Green function $G(x)$ at $\Delta=2$, $\Delta_2=0$ becomes equal to the ground state result for temperatures $T\ll 0.4J$ (the latter being the analytically-known gap).
\bibitem{commentunit}
At finite temperature, the size of the unit cell is doubled. The unit-cell sums in Eqs.~(\ref{eq:gf}), (\ref{eq:sj}), (\ref{eq:cnn}), and (\ref{eq:chi}) are however carried out over the physical sites only.
\bibitem{commentnum}
We test our ground-state numerics by (1) running calculations repeatedly using different (random) initial states, (2) running calculations with successively increased $\chi$, (3) employing different schemes to successively decrease the size of the imaginary time steps,\cite{tebd} (4) varying the accuracy of the ground state energy at which the imaginary time evolution is stopped, (5) comparing the obtained ground state energy for $\Delta_2=0$ with the exact Bethe ansatz result, (6) comparing results for the single-particle Green function $G(x)$ and the density response $C_{NN}(x)$ for $\Delta=\Delta_2=0$ (free fermions) with the exact expression, and (7) checking that the results of a finite-temperature calculation approach the ground state data as $T$ becomes small. All this is necessary to ensure that the imaginary time evolution indeed yields the ground state in the thermodynamic limit.
\end{thebibliography}
\end{document}